\documentclass[
 reprint,
superscriptaddress,
preprintnumbers,
 amsmath,amssymb,
 aps,
prb,
showkeys,
floatfix,
]{revtex4-2}

\usepackage{graphicx}
\usepackage{dcolumn}
\usepackage{subfigure}
\usepackage{bm}
\usepackage[mathlines]{lineno}
\usepackage[utf8]{inputenc}  
\usepackage[unicode=true, bookmarks=false, breaklinks=false, pdfborder={0 0 1},colorlinks=false]{hyperref}
\usepackage{breakurl}
\usepackage[normalem]{ulem} 
\usepackage{url}
\usepackage{gensymb} 
\usepackage{natbib}
\usepackage{multirow}
\usepackage{float}
\usepackage{amssymb}
\usepackage{xcolor}
\usepackage{soul}
\usepackage{amsmath}
\usepackage{mathtools}
\usepackage[T1]{fontenc}
\usepackage{hyperref}

\begin{document}

\preprint{Kurichenko et al./Co-Zr amorphous thin film}

\title{Finite-size effects in amorphous thin Co$_{70}$Zr$_{30}$ layers}
\author{Vladislav Kurichenko}
\affiliation{Department of Physics and Astronomy, Uppsala University, Box 516, SE-751 20 Uppsala, Sweden}
\author{Parul Rani}
\affiliation{Department of Physics and Astronomy, Uppsala University, Box 516, SE-751 20 Uppsala, Sweden}
\affiliation{Institute of Materials Science, Technische\ Universit\"at \ Darmstadt, 64287 \ Darmstadt, Germany}
\author{Björgvin Hjörvarsson}
\email{bjorgvin.hjorvarsson@physics.uu.se}

\affiliation{Department of Physics and Astronomy, Uppsala University, Box 516, SE-751 20 Uppsala, Sweden}

\date{\today}

\begin{abstract}
Profound finite size effects are observed in both the moment and ordering temperature in thin Co$_{70}$Zr$_{30}$ layers.
The results are consistent with the presence of interface regions with reduced magnetic interactions and moment. The extension of this region is determined to be around 1 nm thick at each interface. Above and near the apparent critical temperature, the magnetic properties can be understood in terms of Griffith phases.
\end{abstract}


\maketitle

\section{Introduction}

Amorphous materials are characterized by the absence of long-range structural order. Their properties can therefore become isotropic, as for example observed in the mechanical and magnetic behavior of amorphous alloys. 
Similar to their crystalline counterparts, the physical properties of amorphous alloys are affected by their composition.

For example the saturation magnetization \cite{lohr2021dependence}, magnetic coercivity \cite{xi2019dependence}, as well as the magnetic anisotropy \cite{hebler2016ferrimagnetic}, all depend on the chemical composition. This dependence can be rationalized by the effect of elemental abundance on exchange interactions between neighboring atoms as well as the contribution of each element to the total magnetization of the material. 
For example, increasing the concentration of ferromagnetic elements such as cobalt in Co$_{1-x}$Zr$_{x}$ alloys enhances the saturation magnetization and raises the magnetic ordering temperature \cite{rani2023structural}. Conversely, increasing the concentration of non-magnetic elements tends to decrease the effective magnetic interactions, resulting in a reduction of the magnetic ordering temperature.

Finite-size effects are manifested as changes in physical properties with changes in extension, as $e.g.$ described by Fischer and Barber \cite{fisher1972scaling}.
Numerous models have been developed to describe these effects, ranging from simple approaches that consider only the layer thicknesses and interfaces \cite{xin2014finite} to more sophisticated models that incorporate additional factors, such as magnetic “dead-layer” \cite{kim2001experimental}, shift exponents \cite{huang1993finite} and the effective range of magnetic interactions \cite{taroni2010influence}. Most explorations of finite size of magnetic layers are conducted using single elements. The effects from boundaries are therefore simple to treat and changes in interactions are only taking place in one dimension. Alloying does not affect the material in a homogeneous way, due to the randomness in the configuration of the chemical elements \cite{gemma2020impact}. Therefore, finite-size effects can influence the magnetic behavior of alloys in an unexpected way, due to the spatial dependency of the interactions. 

Well defined finite-size effects have been observed in Fe$_{1-x}$Zr$_{x}$ amorphous alloys \cite{korelis2012finite, ahlberg2012effect, gemma2020impact}, in stark contrast to the results obtained from CoFeTaB layers \cite{tokacc2021magnetic}. The absence of finite size effects in CoFeTaB can be understood by invoking phase separation, as acknowledged by the author. The interface induced phase separation observed in Fe$_{1-x}$Zr$_{x}$ \cite{Korelis2010} might serve as a model of the origin of the suggested phase separation in the CoFeTaB layers. 
Cobalt can, similar to Fe, be made amorphous by the addition of small amounts of zirconium \cite{moubah2013strain}. At low Zr concentrations, the magnetic ordering temperature is only modestly affected, in a stark contrast to Fe$_{1-x}$Zr$_{x}$ alloys. 
The ordering temperature can therefore be varied widely by the choice of composition, which makes the alloy particularly suitable for studying finite-size effects \cite{rani2023structural}. 
Here we use Co$_{70}$Zr$_{30}$ single-layer thin films with an intrinsic ordering temperature (in absence of finite size effects) of at around room temperature, allowing investigations of layers in a wider thickness range without risking annealing effects on the intrinsic magnetic properties of the layers. 

\section{Methods}
Co$_{70}$Zr$_{30}$ single-layer thin films with thicknesses of 25, 50, 75 and 400~\AA\ were deposited at room temperature using ultrahigh-vacuum (UHV) based dc magnetron sputtering. The base pressure of the deposition chamber was $\approx 10^{-10}$~Torr. 
High purity Argon was used as the sputtering gas at a pressure of $\approx 2\cdot10^{-3}$~Torr. The films were grown on a Si(100) substrate mounted on a rotating (6~rpm) holder. Prior to deposition, the substrates were annealed in vacuum at 150~{\textcelsius} for 1~hour. After cooling to ambient temperatures, a 5~nm layer seed layer of Al$_{70}$Zr$_{30}$ was deposited to ensure that the magnetic Co$_{70}$Zr$_{30}$ layer was fully amorphous \cite{Korelis2010}. The same alloy was used as a capping layer to prevent oxidation and to provide symmetric interfaces for the Co$_{70}$Zr$_{30}$ layers. 

The structure and layering of the films were investigated using grazing incidence X-ray diffraction (GIXRD) and X-ray reflectivity (XRR) using a D-8 diffractometer using Cu K$_{\alpha}$ radiation with a wavelength of $\lambda=0.1542$~nm. In GIXRD, the detector angle was varied $2\theta=10^{\circ}-80^{\circ}$ while the incident angle was fixed at $\omega=1.0^{\circ}$. XRR measurements were conducted using $\theta$-$2\theta$ scans up to $2\theta=4.0^{\circ}$.

A Quantum Design MPMS XL superconducting quantum interference device (SQUID) magnetometer was used to measure magnetization of the samples in the temperature range of 5-390~K. The magnetic field was applied in-plane of the sample in all the measurements. The temperature dependence of the magnetization was performed in an applied field of 2~mT. Hysteresis loops were measured at different temperatures (77-300 K), using Magneto-Optical-Kerr-Effect (MOKE) .

\section{Results and discussion}
\subsection{Structure}

\begin{figure}[t]
\includegraphics[width=8cm]{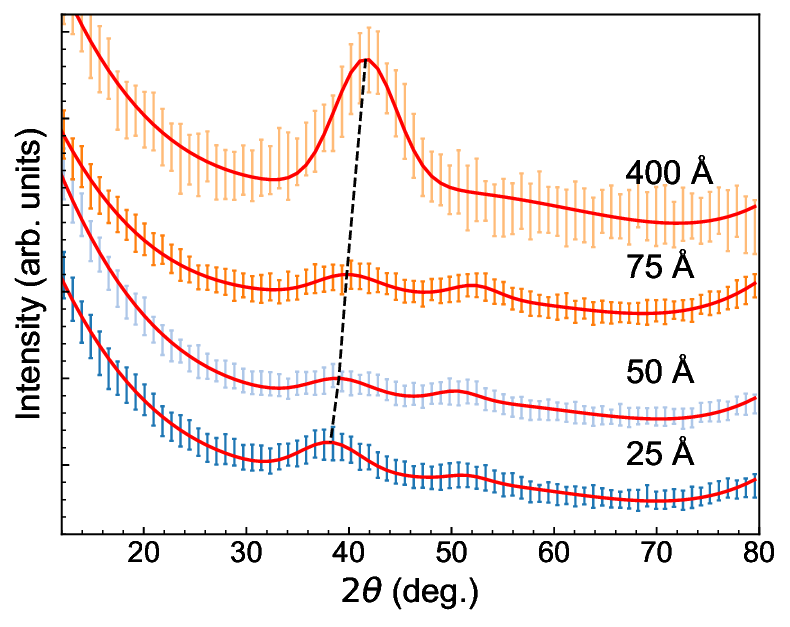}
\caption{\label{fig:XRD}Normalized GIXRD patterns of Co$_{70}$Zr$_{30}$ with different thickness. The patterns were offset for clarity.}
\end{figure}

Results from grazing-incidence X-ray scattering (GIXRD) measurements are shown in Fig.~\ref{fig:XRD}, and the extracted parameters are summarized in Table~\ref{tab:XR}. The 400~\AA-thick sample exhibits a well-defined feature at approximately $2\theta \approx 42^\circ$, with a full width at half maximum (FWHM) of about $6^\circ$, which corresponds to about $ \approx $ 10 ~\AA~ correlation length (Scherrer analysis). In contrast, the 25, 50, and 75~\AA-thick samples display two broad humps, each with a width comparable to that observed in the 400~\AA-thick film. The feature at $2\theta \approx 42^\circ$ has previously been attributed to the Co$_{70}$Zr$_{30}$ layer \cite{rani2023structural}, and the corresponding lower-angle feature observed in the thinner samples is therefore likely to have the same origin. A gradual shift of this feature toward lower angles is observed with decreasing film thickness (dashed black line). This behavior is consistent with the presence of an interfacial region with modified interatomic distances, although additional contributions cannot be excluded. No clear systematic thickness dependence is observed for the hump located just above $2\theta \approx 50^\circ$, which is tentatively assigned to the Al$_{70}$Zr$_{30}$ layers.

Since no direct measurements of spatial variations in the interatomic distances are available, a simple phenomenological model is used to rationalize the observations. We assume that the center position of the scattering feature reflects an average interatomic distance within the Co$_{70}$Zr$_{30}$ layer. The layer is modeled as consisting of an interior region embedded between two interfacial regions characterized by a different average interatomic distance. Within this framework, the effective interatomic distance $d$ can be expressed as
\begin{equation}
\langle d \rangle = d_a \left(1 - \frac{2\Delta}{L}\right) + d_b \frac{2\Delta}{L},
\end{equation}
where $d_a$ and $d_b$ denote the average interatomic distances in the interior and interfacial regions of the Co$_{70}$Zr$_{30}$ layer, respectively. Here, $\Delta$ represents the thickness of a single interfacial region, and $L$ is the thickness of the layer.

When the extracted peak positions are plotted as a function of inverse thickness, the data exhibit an approximately linear dependence in the thickness range from 50 to 400~\AA. Within the context of the model, this behavior suggests that the combined thickness of the interfacial regions, $2\Delta$, lies in the range 25-50~\AA. A fit to the data yields an interior interatomic distance of $d_a = 2.15(1)$~\AA, while the interfacial distance $d_b$ cannot be uniquely determined from the fit alone. However, the product of the interfacial thickness and the difference between $d_a$ and $d_b$ is constrained to be 8.03 (slope of the fit). Based on these constraints, the interatomic distance in the interfacial region is inferred to lie within the range $2.30 < d_b < 2.45$~\AA. Given that the thinnest layer yields an interatomic distance of about 2.35~\AA, and that the thickness is comparable or below the extension of the inferred interface region, the results appear to be plausible. These results are therefore consistent with modified interatomic spacing near the interfaces, although more direct structural probes would be required for a definitive determination.

\begin{figure}[t]
\includegraphics[width=8cm]{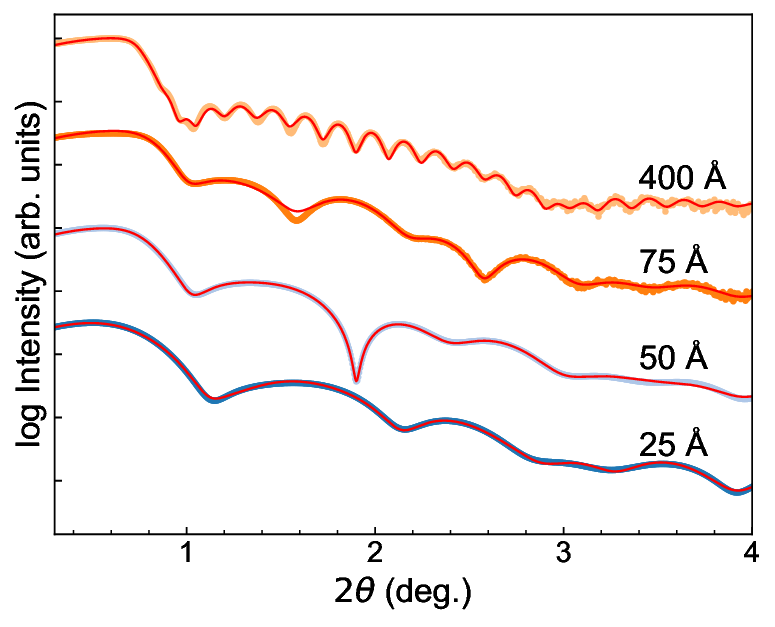}
\caption{\label{fig:XRR}XRR patterns of Co$_{70}$Zr$_{30}$ with different thickness. The solid red lines show the \textit{GenX} \cite{Bjorck2007} fits of the reflectivity data. The patterns were offset for clarity.}
\end{figure}

\begin{table}[t]
\caption{\label{tab:XR}Results of X-ray analysis (XRR and GIXRD). $L_N$ stands for the intended thickness while $L_F$ is the result from the fitting. The peak position and the corresponding interatomic distances are defined in FIG. 2.}
\begin{ruledtabular}
\begin{tabular}{cccccc}
\textrm{$L_N$, \AA}&
\textrm{$L_F$, \AA}&
\textrm{$2\theta_1$, \degree}&
\textrm{$d_1$, \AA}&
\textrm{$2\theta_2$, \degree}&
\textrm{$d_2$, \AA}\cr

\colrule
25 & 28${\pm}$5 & 38.22${\pm}$0.04 & 2.354 & 51.56${\pm}$0.18 & 1.77\\
50 & 48${\pm}$3 & 38.96${\pm}$0.13 & 2.307 & 50.78${\pm}$0.18 & 1.80\\
75 & 73${\pm}$2 & 39.81${\pm}$0.04 & 2.264 & 52.33${\pm}$0.05 & 1.75 \\
400 & 407${\pm}$10 & 41.88${\pm}$0.01 & 2.168 & - & - \\
\end{tabular}
\end{ruledtabular}
\end{table}

The results of the X-ray reflectivity (XRR) measurements are shown in Fig.~\ref{fig:XRR}. The data were analyzed by fitting using the \textit{GenX} software package \cite{Bjorck2007}, and the extracted parameters are summarized in Table~\ref{tab:XR}. The fits indicate well-defined layer thicknesses, with the extracted values ($L_F$) in agreement with the nominal thicknesses ($L_N$) within the experimental uncertainty.

\subsection{Magnetic properties}

\begin{figure}[b]
\includegraphics[width=8cm]{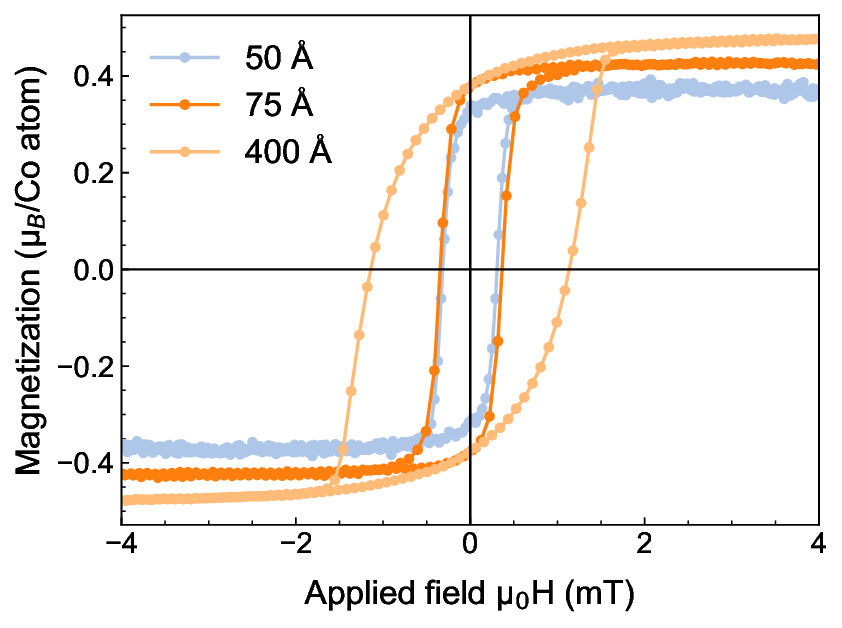}
\caption{\label{fig:MH} M(H) dependence of Co$_{70}$Zr$_{30}$ with different thickness at 100 K obtained using MOKE measurements.}
\end{figure}

Figure~\ref{fig:MH} shows the $M(H)$ dependencies measured at 100~K for three of the samples using MOKE. Data for the 25~\AA-thick film is not shown, as its magnetic ordering temperature lies well below 100~K. As seen in the figure, both the coercivity and the saturation moment are affected by the thickness of the layers. 
When growing random amorphous alloys in a field a well defined uniaxial anisotropy can develop, if the growth temperature is below the magnetic ordering temperature \cite{raanaei2009imprinting}. The origin of the anisotropy is believed to be the interplay between the external field and the evolving moment, affecting the atomic configurations in the growing sample. Consequently, in the absence of an external field, a random local anisotropy can develop when growing layers at temperatures below the magnetic ordering temperature. 

\begin{figure}[t]
\includegraphics[width=8cm]{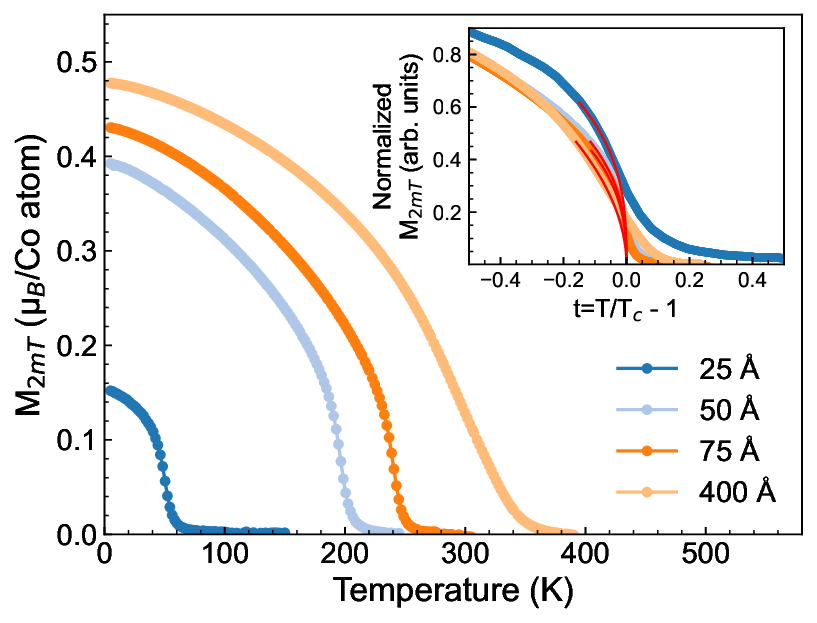}
\caption{\label{fig:MT} M(T) dependence of Co$_{70}$Zr$_{30}$ layers with different thicknesses, determined using SQUID. Inset shows plot of the data in reduced units.}
\end{figure}

The temperature dependence of the magnetization is shown in Fig.~\ref{fig:MT}. The measurements were performed in an external magnetic field of 2~mT, and representative fits to the data are shown in the inset (in reduced units). The ordering temperature of the 400~\AA-film is well above room temperature, which is not the case for the thinner layers, providing plausible reason for the observed differences and similarities in the field response of the samples. The temperature for the onset of ordering as well as the moment per Co atom are also strongly depending on the thickness of the layers, showing a systematic decrease in moment per Co atom with decreasing film thickness. This effect can be ascribed to finite size effects arising from the presence of boundaries in the magnetic layers. As seen in the figure, profound tailing is observed in all the samples. The tailing is partially field induced, while effects from the finite size of the layers will also contribute. An increase in the relative amount of tailing is observed with decreasing Co content \cite{rani2023structural}, reflecting variation in the configurationally depending Co interactions.

In the following, the experimental trends are described using a simple phenomenological model, 
analogous to the description of the structural data above. The magnetic layer is assumed to consist of an interior region and two interfacial regions of thickness $\Delta$. The interface region is assumed to have zero magnetic moment and magnetic interactions. This is the simplest way to describe the effect of interface layers on the resulting magnetic properties, which is the motivation for its use. 
Access to information on the magnetic gradient at the interfaces may motivate the use of a more complex model for the magnetic profile. Under our assumptions, the thickness dependence of the ordering temperature can be written as
\begin{equation}
T_c(L) = T_c(\infty)\left(1 - \frac{2\Delta}{L}\right),
\end{equation}
where $L$ denotes the thickness of the magnetic layers, $\Delta$ is the effective thickness of a single interfacial region, and $T_c(L)$ and $T_c(\infty)$ are the ordering temperatures of the film and the bulk material, respectively \cite{xin2014finite}. In an analogous way, the saturation magnetization can be expressed as,
\begin{equation}
M_s(L) = M_s(\infty)\left(1 - \frac{2\Delta}{L}\right).
\end{equation}

\begin{figure}[t]
\includegraphics[width=8cm]{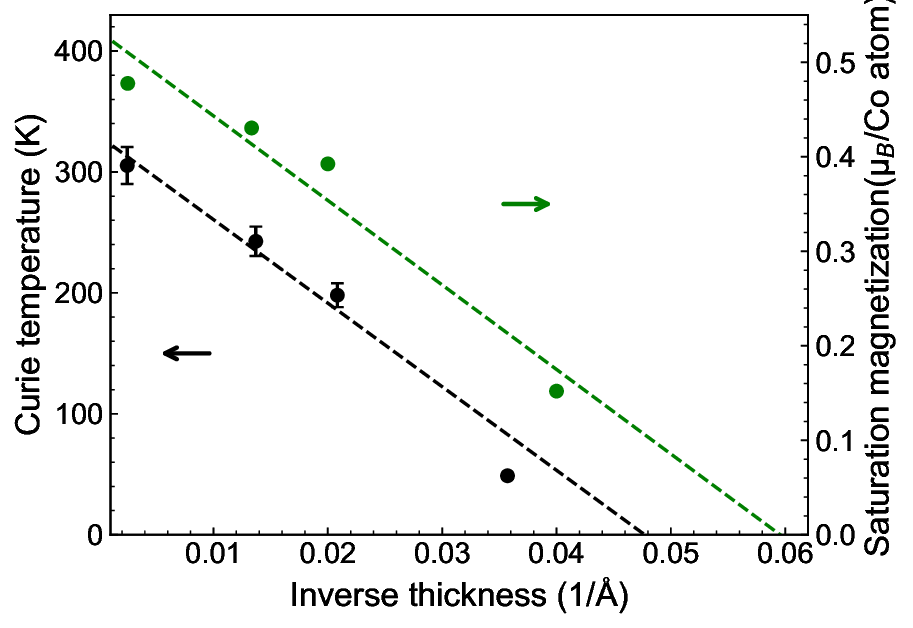}
\caption{\label{fig:Tt_Ms} T$_{C}$(1/L) and M$_{S}$(1/L) dependence of Co$_{70}$Zr$_{30}$. Dashed lines show linear fits to the experimental data.}
\end{figure}

Figure~\ref{fig:Tt_Ms} displays the observed dependence of the magnetic ordering temperature and the saturation magnetization on inverse film thickness. 
Fits to the ordering temperature yields $T_c(\infty)=330$~K which agrees well with the expected value for the given composition \cite{rani2023structural}. 
Using the ordering temperature, the extension of the magnetic interface region is determined to be $\Delta \approx 10$~\AA, which is large as compared to previous results on $\it{e.g.}$ Fe based amorphous \cite{korelis2012finite} and single component \cite{huang1993finite} layers. The extension of the region with affected moment is determined to be $\Delta \approx 8$~\AA.~ The inferred extension of the magnetically dead interface layers is comparable to the deduced extension of the structurally affected interface region. It is therefore not surprising to see the largest deviation from the model in the data obtained from the thinnest layer. 
The saturation magnetization of the interior layer is determined to be $M_s(\infty)=0.53~\mu_B$ from the intercept to the y-axis. 
The magnetic moment of the Co is therefore strongly affected by the alloying as is the effective exchange coupling within the layers. 

The simple model described here provides a reasonable description of the experimental data. However, the model is somewhat oversimplified when $e.g.$ postulating magnetically dead layers, instead of including a gradient in moment and interactions. The effect of inhomogeneous composition at the interfaces can give rise to large proximity effects and non-trivial temperature dependence of the interior and the interface regions \cite{magnus2016long}, which is ignored here. 
We also note that the temperature dependence of the thickness of Fe$_{90}$Zr$_{10}$ layers exhibits much larger deviation from the model \cite{korelis2012finite}. Hence, the composition of the magnetic layers has a strong effect on the apparent thickness dependence on the ordering.
The interface effects in alloys deviate therefore from what is observed in single element layers. This is argued to be at least partially caused by the differences in the boundary between the magnetic and non-magnetic layer, in line with previous suggestions \cite{inyang2023non, gemma2020impact}.

\subsection{A closer look on the transition}

\begin{figure}
    \centering
    \centering
    \centering
    \includegraphics[width=1\linewidth]{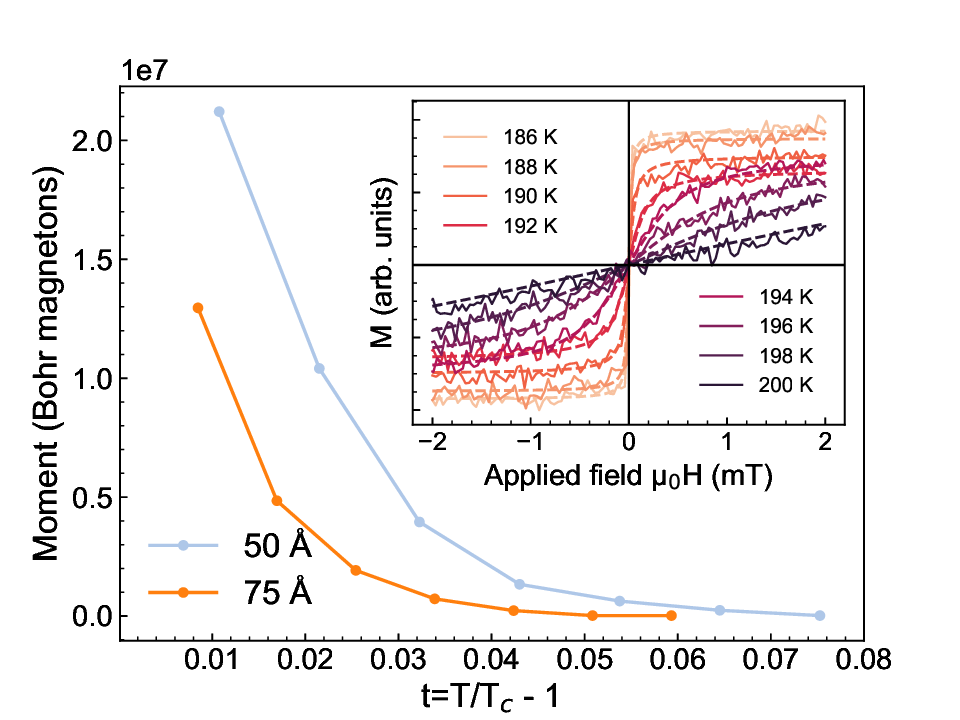}
    \caption{M(t) dependence of Co$_{70}$Zr$_{30}$ with 50~\AA\ and 75~\AA~thickness. The inset shows the fitting of the field dependence of the MOKE results above the ordering temperature for the 50~\AA\ sample. Measurements from the 25~\AA\ sample are not presented as the onset of ferromagnetic order is below the available temperature in the setup (77 K).}
    \label{fig:mu_susc}
\end{figure}

Tailing is well known in the magnetization obtained from two dimensional X-Y magnets, reflecting the presence of correlations well above the ordering temperature \cite{bramwell1993magnetization}. Here the layers have thicknesses which are well beyond the limit of the expected two dimensional spatial limit, implying different origin. 
Above the apparent ordering temperature, both the films exhibit S-shaped $M$–$H$ loops, see inset in Fig.~\ref{fig:mu_susc}, with no remanence. At first sight, hysteresis loops of this type resemble superparamagnetic state arising from finite-sized ferromagnetic clusters, in contrast to long-range magnetic order \cite{liebig2011experimental}. The MOKE results were calibrated using the SQUID data and the magnetization was scaled by the film thickness. The results were fitted using a Langevin function,
\begin{equation}
M(H) = M_s L(\mu H / k_{\rm B}T).
\end{equation}
providing information on the length scale of the magnetic fluctuations. 
Here $L(x)$ is the Langevin function and $\mu$ is the magnetic moment of the dynamic macrospins. The results are shown in Fig.~\ref{fig:mu_susc}. Representative fitting of the field dependence obtained from one of the samples, is shown as an inset in the figure. 
An increase in size of the dynamic moment is observed with decreasing thickness of the layers, well above the transition temperature. For example, at $t=0.02$, about twice as large dynamic moments are obtained in the thinner layers. Considering that the moment originates in a collective response of Co, an increase in the relative correlation length is obtained with decreasing thickness of the layers. 
The magnetic susceptibility provides additional insight, while the response for the 75~\AA\ sample has relatively small width (few K) the 50~\AA\ film exhibits much broader and ill defined response in temperature. The results are therefore consistent with a transition of a non-homogeneous sheet, shifting the effective magnetic percolation limit when changing the thickness of the layers \cite{vojta2006rare, gemma2020impact}.

The obtained magnetic behavior can be understood as transition from a ferromagnetic to a Griffiths phase \cite{griffiths1969nonanalytic, bray1987nature}. In a randomly disordered magnetic material there will be variations in the effective coupling strength between the atoms carrying a magnetic moment. Consequently, there will be regions with locally enhanced exchange interactions (and thus higher local "Curie" temperatures $T_C^{\rm local}$). Those regions can remain internally ordered, well above the apparent ordering temperature of the material. Hence, the effect is similar as having isotropic ferromagnetic clusters embedded in an paramagnetic matrix \cite{chan2006critical}.
Here, these dynamic clusters give rise to an enhanced susceptibility and magnetization above the nominal $T_C$ for highly disordered 50~\AA-thick sample. In contrast, the thicker 75~\AA\ film, with more bulk-like exchange coupling, exhibits fewer or smaller such clusters above $T_C$, resulting in smaller $\mu$ and more rapidly vanishing $\chi(T)$ above the critical temperature, separating the ferromagnetic and the Griffiths phases.

\section{Conclusions}
The investigations showed that changes in both saturation magnetization and apparent ordering temperature can be understood using a simple interface model, where both the ordering temperature and moment scale with the inverse thickness of the magnetic layers. In the thin film limit, the description halts and the lateral variations in the effective coupling needs to be included to capture the observed field dependence above the apparent ordering temperature. In this limit the ordering temperature is expected to scale with the coverage of a critical concentration, giving rise to linear dependence of the ordering temperature as observed by \textit{Ahlberg et. al.} \cite{ahlberg2011two}.

The behavior highlights the importance of compositional and structural disorder in alloys on magnetic phase transitions. These effects are argued to increase in importance when the extension of the material becomes comparable to the length scale of the density variations in composition \cite{gemma2020impact}. Even in a nominally homogeneous alloys, variations in exchange coupling and local atomic environments can lead to Griffiths-phases, masking the ferromagnetic to paramagnetic phase transition. These results are argued to arise from the presence of local order, caused by areal densities of the magnetic layers being below the percolation limit with respect to ferromagnetic exchange.

\section*{Acknowledgments}
This work was performed with a partial financial support from the Swedish Research Council (Grant No. 2017-03725).
P.R. would like to acknowledge Prof. Gabriella Andersson and Dr. Petra Erika J\"{o}nsson for their support and supervision during the study, as well as for their assistance with the MOKE measurements.

\bibliography{apssamp}

\end{document}